%Paper: hep-th/9308014
%From: donatis@tsmi19.sissa.it
%Date: Thu, 05 Aug 1993 11:42:15 +0100

\font\smc = cmcsc10

\magnification=1200

\topskip=25pt

\baselineskip=14pt
\hfuzz=20pt

\def\p{\par\noindent}
\def\v{v^2}
\def\ie{{\it i.e.}\ }
\def\vp{\bigl(|\phi|^2-\v\bigr)}
\def\D{\vec D}
\def\A{\vec A}
\def\dps{|\D\phi|^2}
\def\II{\int d^2r\,}
\def\At{A_\theta}
\def\DD{\bar D}
\def\pa{\partial}

\def\pt{$P$($T$)\ }
\def\bcs{B^{CS}}
\def\J{\vec J}
\def\ACS{\A^{CS}}

\font \titlefont=cmr10 scaled\magstep1

\null
\vskip-2truecm
\rightline{{\bf  S.I.S.S.A. 129/93/EP }}
\vskip2truecm
\vskip0.5truecm
\centerline{\titlefont  COMMENT ON VORTICES IN CHERN-SIMONS}
\centerline{\titlefont  AND MAXWELL ELECTRODYNAMICS}

\vskip3truecm
\centerline{Pietro Donatis and Roberto Iengo}
\vskip 1.0truecm
\centerline{{\it International School for Advanced Studies,
                I-34014 Trieste, Italy and}}
\centerline{{\it Istituto Nazionale di Fisica
                Nucleare, INFN, Sezione di Trieste, Trieste, Italy}}

\vskip3.0truecm
\noindent{\bf Abstract}
\vskip.3truecm
\par
\noindent
We compare the vortex-like solutions of two different theories in ($2+1$)
dimensions. In the first a nonrelativistic field self-interacts through a
Chern-Simons gauge connection. It is $P$ and $T$ violating. The second is
the standard Maxwell scalar electrodynamics. We show that for specific
values of some parameters the same vortex-configurations provide solutions
for both theories.
\vfill
\eject

\newcount\equanumber	\equanumber=0
\def\chapterlabel{1}

\def\eqname#1{\relax\global\advance\equanumber by 1
  \xdef#1{{\rm(\chapterlabel.\number\equanumber)}}#1}

\def\eqn{\eqno\eqname}

\newcount\refnumber	\refnumber=0
\def\refname#1{\relax\global\advance\refnumber by 1
  \xdef#1{{\rm[\number\refnumber]}}#1}
\def\ref{\refname}

\centerline{\bf {1. Introduction}}
\bigskip
\noindent
Vortices in Field Theory have been much studied in the literature (let us
quote for example the pioneering article by H.B. Nielsen and P. Olesen
\ref\Nielsen). An isolated vortex is basically a configuration in two space
dimensions and thus represents a typically time independent extended object
of a ($2\!+\!1$) space-time dimensional theory. It is also well known that
in this dimensionality a gauge theory allows for the interesting
possibility of a Chern Simons ($CS$) gauge field (for some general
introduction see \ref\Templeton\ and \ref\Deser). And indeed the vortices
has been also studied in the $CS$ case, both relativistic and
nonrelativistic (see for instance references
\ref\Vega\ \ref\Schaposnik\ \ref\Lozano\ \ref\Weinberg\ \ref\Paul\ \ref\Hong).
\p
In this letter we present an additional theoretical observation in this
topic, by comparing the vortex-like solutions of a ($2\!+\!1$)-dimensional
Maxwell electrodynamics coupled to nonrelativistic matter and the
vortex-like solutions of a different theory in which the gauge field is a
($CS$) one. Though this letter is ``self-contained'' we refer to \ref\Io\
for a more detailed discussion on the physical properties of this $CS$
theory.
\p
In particular we will study two different Hamiltonians: the first is the
$CS$ one:
$$
H=\II\biggl\{{1\over 2}\dps+g\vp^2\biggr\}
\eqn\hchiral
$$
\p
In \hchiral\ $\phi$ is a nonrelativistic complex field,
$\D\!=\!\vec\nabla\!-\!i\ACS$, where $\ACS$ is given by the assumption that
its field strength is determined by the matter density fluctuation from the
mean value $\v\!=<\!|\phi|^2\!>$, that is:
$$
\bcs=\vec\nabla\wedge\ACS= {2\pi\over k}\vp
\eqn\fabio
$$
here $k$ is a constant, which we take to be positive.
\p
Differently from \Io, here we take the value of the mass $m$ to be $1$,
which is possible through a rescaling of $\phi$, $v$, $g$ and $k$.
\p
The motivation for assuming equation \fabio, rather than the assumption
that the $CS$ field strength is proportional to the matter density, is
discussed in \Io. Here we remark that equation \fabio\ makes consistent the
formulation since by taking a fixed total number of particles $N$, as
appropriate in a non relativistic theory, we have consistently:
$$
\II\vec\nabla\wedge\ACS=0
\eqn\
$$
This theory is formally the same as the one introduced in references
\ref\ZHK\ \ref\Lee\ \ref\Fisher\ \ref\Zhang, where also vortices has been
studied, for describing the Fractional Quantum Hall Effect. Here we focus
on some field theoretical aspects of the problem, and thus we will not be
concerned with particular applications.
\p
The second Hamiltonian that we will study is the Maxwell one:
$$
H=\II\biggl\{{1\over 4} F_{ij}F_{ij}+{1\over 2}\dps+g\vp^2\biggr\}
\eqn\lagr
$$
In equation \lagr\ the covariant derivative $\D\!=\!\vec\nabla\!-\!i\A$
contains the Maxwell gauge field $\A$ and $F_{ij}$ is the corresponding
field strength, determined by the usual Maxwell equation:
$$
\pa_j F_{ji}=-J_i
\eqn\
$$
where $\J$ is the current
$\J\!=\!{1\over 2i}\Bigl[\phi^{\dag}\D\phi-\phi(\D\phi)^{\dag}\Bigl]$.
\p
The first theory is not invariant under parity ($P$) and time reversal ($T$)
transformations (for this reason we will call it {\it chiral} theory),
while the second one is $P$ and $T$ invariant.
\p
We are interested in the vortex solutions of these two theories.
\p
Let us parameterize the vortex configuration of the matter field $\phi$ by:
$$
\phi(r,\theta)=f(r)e^{in\theta} \qquad \lim_{r\to\infty}f(r)=v
\eqn\vortex
$$
here $r$ and $\theta$ are circular polar coordinates.
\p
Then the set of vortex solutions falls in ``topological classes'' labelled
by the integer $n$ which is a topological invariant called {\it vorticity}.
This integer corresponds, in a rough sense, to how many times the vortex
winds round.
\p
In two dimensions the $P$ transformation corresponds to the reflection of
only one axis. Here we will take the case:
$$
\left\{
\eqalign{&x\to -x\cr
&y\to y\cr}\right.
\eqn\
$$
which in polar coordinate reads:
$$
\left\{
\eqalign{&r\to r\cr
&\theta\to\pi-\theta\cr}\right.
\eqn\parity
$$
On the other hand  the $T$ transformation changes the sign of the time
coordinate and takes the complex conjugation of the scalar fields. Actually
in our considerations the time coordinate plays no r\^ole. But both for $P$
and $T$ transformations the vortex \vortex\ transforms as:
$$
\phi(r,\theta)=f(r)e^{in\theta}\to f(r)e^{-in\theta}
\eqn\
$$
apart for a possible inessential overall sign, \ie $P$ and $T$ connect the
topological classes that differ by the sign of the vorticity. Therefore we
expect that in a chiral theory vortex solutions that differ by the sign of
the vorticity will be inequivalent. Indeed we have found in \Io\ that they
have different energy.
\p
In Maxwell electrodynamics we instead expect a perfect symmetry of the
vortex solutions that differ only by the sign of $n$.
\p
In this letter we will show that the chiral theory \hchiral\ and the Maxwell
theory \lagr\ admit as solutions, though for different values of the
constant $g$, exactly the same vortex configurations. Furthermore we will
show that the vortex configurations that are solutions of two
{\it different} chiral theories connected by a \pt transformation differ
only by the sign of the vorticity, and are both solutions of the Maxwell
theory.
\p
In the second section we will study the chiral theory discussing separately
the two cases of $n\!<\!0$ and of $n\!>\!0$. Then, in the third section, we
will turn to the Maxwell electrodynamics.
\goodbreak
\bigskip
\bigskip
\def\chapterlabel{2}
\newcount\equanumber	\equanumber=0
\centerline{\bf{2. Chiral theory}}
\nobreak
\bigskip
\noindent
{\it i) case $n\!<\!0$}
\medskip
\noindent
Let us now start from the Hamiltonian \hchiral\ and study it following the
method of refs. \Paul\ \Hong\ \ref\Jack\ based on the classical
work of \ref\Bogo.
\p
Using the following identity \ref\Pi:
$$
|\D\phi|^2=\bigl|(D_x\pm iD_y)\phi\bigr|^2
\pm \vec\nabla\wedge\J\pm \bcs |\phi|^2
\eqn\identity
$$
the Hamiltonian \hchiral\ can be written as:
$$
H=\II\biggl\{{1\over 2}\bigl|(D_x\pm iD_y)\phi\bigr|^2\pm
{1\over 2}\vec\nabla\wedge\Bigl(\vec J+\v\ACS\Bigr)+
\Bigl(g\pm {\pi\over k}\Bigr)\vp^2\biggr\}
\eqn\marta
$$
The contribution of the second term can be computed using the fact that
the asymptotic behaviour of the gauge field is:
$$
\lim_{r\to\infty}\At (r)={n\over r}\to 0
\eqn\
$$
this equation comes from the requirement of finite energy and leads to the
quantization of the magnetic flux in integers factors of $2\pi$. So we get:
$$
H=\pm\pi\v n+
\II\biggl\{{1\over 2}\bigl|(D_x\pm iD_y)\phi\bigr|^2
+\Bigl(g\pm {\pi\over k}\Bigr)\vp^2\biggr\}
\eqn\nolasco
$$
Let us consider the special case $g\!=\!{\pi\over k}$\footnote{$^{(1)}$}
{If one wants to relate \hchiral\ to the Anyon mean field theory one has to
take $g\!\ge\!{\pi\over k}$. For a detailed discussion on this, see
\Io\ where the relation between the value of $g$ and the properties of the
spectrum is also discussed.}, $n\!<\!0$. In this case the problem of
minimizing this Hamiltonian greatly simplifies since, taking the lower
(minus) sign in \nolasco, we get:
$$
H=\pi\v |n|+
{1\over 2}\II\bigl|(D_x-iD_y)\phi\bigr|^2
\eqn\andrea
$$
This expression is minimal for:
$$
(D_x-iD_y)\phi=0
\eqn\self
$$
which corresponds to the energy:
$$
E_n=\pi\v |n|
\eqn\minimal
$$
In polar coordinates equation \self\ reads:
$$
\pa_rf+{n\over r}f-\At f=0
\eqn\zaffa
$$
if we introduce the auxiliary variable:
$$
a(r)=-n+r\At(r)
\eqn\
$$
with the properties: $a(0)\!=\!-n\!>\!0$, $a(\infty)\!=\!0$, we can
rewrite \zaffa\ as:
$$
\left\{
\eqalign{&\pa_rf={1\over r}af\cr
&\pa_ra={2\pi\over k}(f^2-\v)\cr}\right.
\eqn\system
$$
which has been solved numerically, see \Io.
\p
The same Hamiltonian \hchiral, with the same constraint \fabio, also admits
vortices with $n\!>\!0$. They have been studied in ref. \Io. However their
energy is not minimal like \minimal\ and they will not be solutions of a
first order equation like \self, or its complex conjugate.
\bigskip
\noindent
{\it ii) case $n\!>\!0$}
\medskip
\noindent
Consider now a {\it different} theory obtained from the one studied so far
by a \pt transformation.
\p
In this new theory, $\bcs$ changes sign so we have:
$$
\bcs=\vec\nabla\wedge\ACS=-{2\pi\over k}\vp
\eqn\posfabio
$$
so in the place of \marta\ now we have:
$$
H=\II\biggl\{{1\over 2}\bigl|(D_x\pm iD_y)\phi\bigr|^2\pm
{1\over 2}\vec\nabla\wedge\Bigl(\vec J+\v\ACS\Bigr)+
\Bigl(g\mp {\pi\over k}\Bigr)\vp^2\biggr\}=
$$
\nobreak
$$
=\pm\pi\v n+
\II\biggl\{{1\over 2}\bigl|(D_x\pm iD_y)\phi\bigr|^2
+\Bigl(g\mp {\pi\over k}\Bigr)\vp^2\biggr\}
\eqn\sciama
$$
With the same value of $g$, taking $n\!>\!0$ and choosing the higher sign,
equation \sciama\ becomes:
$$
H=\pi\v n+
{1\over 2}\II\bigl|(D_x+iD_y)\phi\bigr|^2
\eqn\
$$
This is minimal if the following equation is satisfied:
$$
(D_x+iD_y)\phi=0
\eqn\antiself
$$
which can be obtained from \self\ by a \pt transformation, and corresponds
to an energy equal to \minimal.
\p
In polar coordinates \antiself\ reads:
$$
\pa_rf-{n\over r}f+\At f=0
\eqn\polar
$$
Introducing the auxiliary variable:
$$
b(r)=n-r\At(r)
\eqn\
$$
with the properties: $b(0)\!=\!n\!>\!0$, $b(\infty)\!=\!0$ we get exactly the
same equations as \system:
$$
\left\{
\eqalign{&\pa_rf={1\over r}bf\cr
&\pa_rb={2\pi\over k}(f^2-\v)\cr}\right.
\eqn\
$$
So we have checked that two chiral theories connected by a \pt
transformation admit the same vortex solutions with $n\!\to\!-n$.
\p
We will show next that the \pt invariant Maxwell electrodynamics theory
possesses both kinds of solutions with $n\!<\!0$ and with $n\!>\!0$, where
in both cases the field $\phi$ satisfies a first order equation like
\self\ or \antiself.
\goodbreak
\bigskip
\bigskip
\def\chapterlabel{3}
\newcount\equanumber	\equanumber=0
\centerline{\bf {Maxwell electrodynamics}}
\bigskip
\nobreak
\noindent
Let us start from the Hamiltonian \lagr\ and let us look for its
vortex-like configurations of minimal energy.
\p
Using again the identity \identity\ (now with
$\bcs\to B\!=\!\vec\nabla\wedge\A$) equation \lagr\ can be
rewritten as:
$$
H=\II\biggl\{{1\over 4} F_{ij}F_{ij}+
{1\over 2}\bigl|(D_x\pm iD_y)\phi\bigr|^2\pm
{1\over 2}\vec\nabla\wedge\vec J\pm
{1\over 2}B|\phi|^2+g\vp^2\biggr\}
\eqn\relagr
$$
\bigskip
\noindent
{\it i) case $n\!<\!0$}
\medskip
\noindent
We start discussing the case with the lower sign. Integrating the second
term of \relagr\ by parts we get:
$$
H=\II\biggl\{{1\over 4} F_{ij}F_{ij}-2\phi^{\dag}\DD D\phi-
{1\over 2}\vec\nabla\wedge\vec J-
{1\over 2}B|\phi|^2+g\vp^2\biggr\}
\eqn\comp
$$
In equation \comp\ we have introduced the complex notation for the
covariant derivatives:
$$
D={1\over 2}\bigl(D_x-iD_y\bigr) \qquad \DD={1\over 2}\bigl(D_x+iD_y\bigr)
\eqn\
$$
The equations of motion from this Hamiltonian are:
$$
{\delta H\over\delta\phi^{\dag}}=0 \qquad \Rightarrow \qquad
-2\DD D\phi-{1\over 2}B\phi+2g\vp\phi=0
\eqn\rellia
$$
$$
{\delta H\over\delta A_i}=0 \qquad \Rightarrow \qquad
\pa_j F_{ji}=-J_i
\eqn\rellib
$$
Notice that equation \rellia\ admits the solution:
$$
D\phi=0
\eqn\motiona
$$
$$
B={2\pi\over k}\vp
\eqn\motionb
$$
provided we take $g\!=\!{\pi\over 2k}$, and \rellib\ is satisfied if $B$
is equal to \motionb\ with $k\!=\!4\pi$ as can be easily checked.
Indeed equation \rellib\ in polar coordinates reads:
$$
\left\{
\eqalign{&\pa_rB=-J_{\theta}\cr
&{1\over r}\pa_{\theta}B=J_r\cr}\right.
\eqn\grieg
$$
and computing $J_{\theta}$ and $J_r$ for the vortex
$\phi(r,\theta)\!=\!f(r)e^{in\theta}$ we get:
$$
\left\{
\eqalign{&J_{\theta}=f(r)\Bigl[{n\over r}-\At (r)\Bigr]f(r)\cr
&J_r=0\cr}\right.
\eqn\luca
$$
Now, using equation \zaffa, the first of \luca\ becomes:
$$
J_{\theta}=-f(r)\pa_rf(r)
\eqn\
$$
On the other hand from \motionb\ we get:
$$
\pa_rB={4\pi\over k}f(r)\pa_rf(r)
\eqn\
$$
proving the validity of the first of \grieg. The second is obviously true.
\p
Formally \motiona\ and \motionb\ are exactly the equations for $\phi$ and
$B$ found in the chiral theory described above (see equations \fabio\ and
\self).
\p
Notice that, because of the presence of the kinetic term for the gauge
field, in the Maxwell case $g$ has no more the value ${\pi\over k}$ it had
in the chiral case. Furthermore, now the value of the constant $k$ is no longer
free but it is constrained.
\p
The Hamiltonian \lagr\ can be written as:
$$
H=\II\biggl\{{1\over 2}B^2+{1\over 2}\dps+g\vp^2\biggr\}
\eqn\ham
$$
using \motionb, and \identity, equation \ham\ can be rewritten as:
$$
H=\II\biggl\{2|D\phi|^2-
{1\over 2}\vec\nabla\wedge\Bigl(\vec J+\v\A\Bigr)+
\Bigl(g+{2\pi^2\over k^2}-{\pi\over k}\Bigr)\vp^2\biggr\}
\eqn\
$$
The second term can be computed again using the asymptotic behaviour of
$\A$. The third term vanishes using the values of $g$ and of $k$ found
above, so we get:
$$
H=-\pi\v n+2\II |D\phi|^2
\eqn\urca
$$
Notice that, since for our solution \motiona\ holds, the positivity of
$H$ requires $n$ to be negative.
\p
Formally, equation \urca\ is equal to the corresponding Hamiltonian of the
chiral theory \andrea.
\goodbreak
\bigskip
\noindent
{\it ii) case $n\!>\!0$}
\nobreak
\medskip
\noindent
Now let us discuss the case in which the identity \identity\ is used with
the other (upper) sign. The Hamiltonian \lagr\ can be rewritten in the
following way:
$$
H=\II\biggl\{{1\over 4} F_{ij}F_{ij}-2\phi^{\dag}D \DD\phi+
{1\over 2}\vec\nabla\wedge\vec J+
{1\over 2}B|\phi|^2+g\vp^2\biggr\}
\eqn\hamipos
$$
We stress that this is the {\it same} Hamiltonian as equation \comp.
\p
Again we can compute the equations of motion:
$$
{\delta H\over\delta\phi^{\dag}}=0 \qquad \Rightarrow \qquad
-2D\DD\phi+{1\over 2}B\phi+2g\vp\phi=0
\eqn\
$$
$$
{\delta H\over\delta A_i}=0 \qquad \Rightarrow \qquad
\pa_j F_{ji}=-J_i
\eqn\
$$
These equations are solved by:
$$
\DD\phi=0
\eqn\motionposa
$$
\nobreak
$$
B=-{2\pi\over k}\vp
\eqn\motionposb
$$
with the previous values of $g$ and $k$.
\p
As noted in the chiral case these are the equations that are obtained from
\motiona\ and \motionb\ using the parity transformation \parity.
\p
Equations \motionposa\ and \motionposb\ are again formally equal to the
corresponding equations for $\phi$ and $B$ in the chiral theory (see equations
\posfabio\ and \antiself), so they have the same vortex solutions.
\p
The Hamiltonian \hamipos\ can be rewritten as
$$
H=\pi\v n+2\II |\DD\phi|^2
\eqn\elena
$$
Now the positivity of $H$ requires $n\!>\!0$.
\p
Again equation \elena\ is formally equal to the corresponding Hamiltonian
of the chiral theory \sciama.
\p
Concluding we have obtained exactly the same vortex solutions found in the
chiral theories, as announced in the introduction.

\goodbreak
\bigskip
\bigskip
\bigskip
\centerline{\bf{References}}
\nobreak
\bigskip

\noindent
\Nielsen\ {\smc H.B. Nielsen, P. Olesen},
{\it Nucl. Phys.} {\bf B61} (1973), 45.

\noindent
\Templeton\ {\smc R. Jackiw, S. Templeton},
{\it Phys. Rev.} {\bf D23} (1981), 2291.

\noindent
\Deser\ {\smc S. Deser, R. Jackiw, S. Templeton},
{\it Ann. Phys.} {\bf 140} (1982), 372.

\noindent
\Vega1\ {\smc H.J. Vega, F.A. Schaposnik},
{\it Phys. Rev.} {\bf D14} (1976), 1100.

\noindent
\Schaposnik\ {\smc H.J. Vega, F.A. Schaposnik},
{\it Phys. Rev.} {\bf D34} (1986), 3206.

\noindent
\Lozano\ {\smc G. Lozano, M.V. Manies, F.A. Schaposnik},
{\it Phys. Rev. Lett.} {\bf 56} (1986), 2564.

\noindent
\Weinberg\ {\smc R. Jackiw, K. Lee, E.J. Weinberg},
{\it Phys. Rev.} {\bf D42} (1990), 3491.

\noindent
\Paul\ {\smc S.K. Paul, A. Khare},
{\it Phys. Lett.} {\bf B174} (1986), 420.

\noindent
\Hong\ {\smc J. Hong, Y. Kim, P.Y. Pac},
{\it Phys. Rev. Lett.} {\bf 64} (1990), 2230.

\noindent
\Io\ {\smc P. Donatis, R. Iengo}, Preprint SISSA 101/93/EP,
cond-mat/9307013.

\noindent
\ZHK\ {\smc S.C. Zhang, T.H. Hansson, S.Kivelson},
{\it Phys. Rev. Lett.} {\bf 62} (1989), 82.

\noindent
\Lee\ {\smc D.H. Lee, S.C. Zhang},
{\it Phys. Rev. Lett.} {\bf 66} (1991), 1220.

\noindent
\Fisher\ {\smc D.H. Lee, M.P.A. Fisher},
{\it Int. J. Mod. Phys.} {\bf B5} (1991), 2675.

\noindent
\Zhang\ {\smc S.C. Zhang},
{\it Int. J. Mod. Phys.} {\bf B6} (1992), 25.

\noindent
\Jack\ {\smc R. Jackiw, E.J. Weinberg},
{\it Phys. Rev. Lett.} {\bf 64} (1990), 2234.

\noindent
\Bogo\ {\smc E.B. Bogomol'nyi},
{\it Sov. J. Nucl. Phys.} {\bf 24} (1976), 449.

\noindent
\Pi\ {\smc R. Jackiw, S.Y. Pi},
{\it Phys. Rev. Lett.} {\bf 64} (1990), 2969.

\vfill
\eject
\bye